\documentclass[amsmath,floatfix,twocolumn,superscriptaddress]{revtex4-1}
\usepackage{amssymb,amsmath}
\usepackage{graphicx,array}
\usepackage{dcolumn}
\usepackage{psfrag}
\usepackage{bm}
\usepackage{color}

\begin{document}
\title{Oxygen-vacancy induced magnetic phase transitions in multiferroic thin films}

\author{Cesar Men\'{e}ndez}
\affiliation{School of Materials Science and Engineering, UNSW Sydney, NSW 2052, Australia} 

\author{Dewei Chu}
\affiliation{School of Materials Science and Engineering, UNSW Sydney, NSW 2052, Australia}

\author{Claudio Cazorla}
\thanks{Corresponding Author}
\affiliation{School of Materials Science and Engineering, UNSW Sydney, NSW 2052, Australia}

\maketitle

{\bf Multiferroics in which giant ferroelectric polarization and magnetism coexist are 
of tremendous potential for engineering disruptive applications in information storage 
and energy conversion. Yet the functional properties of multiferroics are thought to be 
affected detrimentally by the presence of point defects, which may be abundant due to the 
volatile nature of some constituent atoms and high temperatures involved in materials preparation. 
Here, we demonstrate with theoretical methods that oxygen vacancies may enhance the functionality 
of multiferroics by radically changing their magnetic interactions in thin films. Specifically, 
oxygen vacancies may restore missing magnetic super-exchange interactions in large axial ratio 
phases, leading to full antiferromagnetic spin ordering, and induce the stabilization of 
ferrimagnetic states with a significant net magnetization of $0.5$~$\mu_{B}$ per formula unit. 
Our theoretical study should help to clarify the origins of long-standing controversies in 
bismuth ferrite and improve the design of technological applications based on multiferroics.}
\\

\begin{figure*}
\centerline
        {\includegraphics[width=1.00\linewidth]{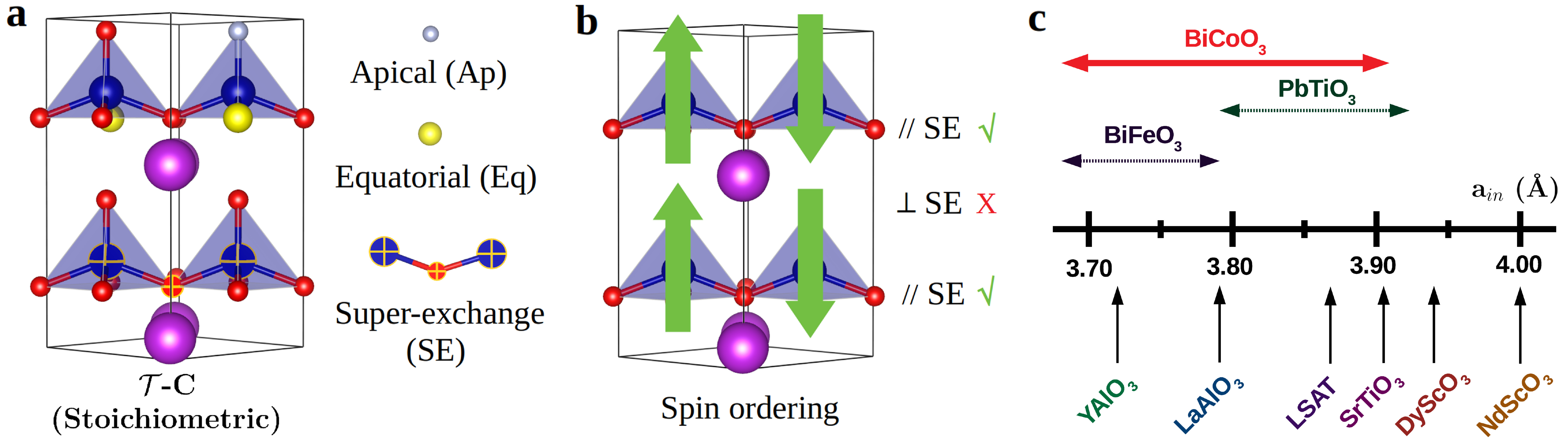}}
	\caption{Sketch of multiferroic BCO in bulk and thin film geometries. {\bf a} Representation 
	of the super-tetragonal phase (${\cal T}$, space group $P4mm$) characteristic of bulk 
	BCO and other multiferroics exhibiting giant electric polarization. {\bf b} 
	Magnetic structure in the ${\cal T}$ phase rendering C-type antiferromagnetism (AFM-C); green
	arrows represent atomic magnetic moments and their orientation. {\bf c} Examples of substrates 
	in which to grow BCO thin films displaying the effects predicted in this study; other well-known
	materials exhibiting super-tetragonal phases are shown for comparison \cite{zhang18,wang03}. 
	``SE'' stands for magnetic super-exchange interactions. Bi, Co, and O atoms are represented with 
	magenta, blue, and red spheres, respectively.}
\label{fig1}
\end{figure*}

Finding multiferroics in which ferroelectricity and magnetism coexist and influence each other 
is of great fundamental and applied interests \cite{spalding07,spalding19}. Salient technological 
features of multiferroics include the possibility of controlling the magnetization with electric 
fields to design efficient logic and memory devices \cite{heron14,allibe12}, and of realizing large 
piezomagnetic coefficients to facilitate the miniaturization of antennas and sensors \cite{domann17,nan17}. 
Furthermore, competition between phases displaying distinct electric polarization and magnetic ordering
offers also encouraging prospects for energy conversion applications like photovoltaics and 
solid-state cooling \cite{huang17,stern18,cazorla18}. 

Unfortunately, multiferroics are rare in nature, typically present weak magnetoelectric coupling 
(BiFeO$_{3}$) \cite{bertinshaw16}, and require extreme synthesis conditions (PbVO$_{3}$ and BiCoO$_{3}$) 
\cite{belik05,belik06}. In addition, magnetoelectric multiferroics mostly are antiferromagnetic hence 
potential applications based on external magnetic bias are frustrated due to the little effect on antiparallel 
magnetic spins \cite{wang18}. Common strategies employed to synthesize bettered multiferroic materials 
include doping \cite{das12,das16}, solid solutions \cite{sakai11,hojo18}, and strain engineering in thin 
films \cite{spalding07,martin08}.

Through epitaxial strain is actually possible to create new multiferroic materials in the laboratory 
that exhibit giant electric polarization and unexpected magnetic spin ordering \cite{zhang18,varga17,goodenough97}. 
An illustrative example is given by BiFeO$_{3}$ (BFO), in which large spontaneous polarization and 
ferromagnetism (FM) have been observed under moderate compressive biaxial strains at room temperature \cite{wang03,scott05}. 
The origins of the net magnetization in BFO thin films, however, are not clear yet and from 
a technological point of view is crucially important to understand them at the fundamental level. 

Most magnetic ferroelectrics with chemical formula A$M$O$_{3}$ and perovskite-like structure present 
antiferromagnetic spin ordering along the three pseudo-Cartesian directions (AFM-G), due to the dominant 
role of oxygen-mediated super-exchange interactions between neighbouring transition metal atoms $M$ 
\cite{goodenough55,filipetti02,spaldin-book,may14}. In large axial ratio structures, the covalency 
of $M$-O bonds parallel to the electric polarization is significantly reduced and consequently 
magnetic exchange interactions, favouring parallel magnetic spins, dominate in that direction; the 
coexistence of ``in-plane'' antiferromagnetism and ``out-of-plane'' ferromagnetism leads anyway to 
null crystal magnetization (AFM-C, Fig.\ref{fig1}a,b) when small spin canting effects are neglected. 
Therefore, intrinsic and robust FM in principle is not expected to occur in BFO or any other similar 
multiferroic \cite{cazorla13,cazorla17,singh06,solovyev12}. 

A plausible explanation for the appearance of FM in A$M$O$_{3}$ perovskite oxide thin films is based 
on extrinsic causes like point defects \cite{scott05,niu18}. The volatile nature of bismuth and 
high temperatures involved in the preparation of samples, for instance, make the presence 
of oxygen vacancies ($V_{O}$) almost inevitable in Bi-based multiferroics \cite{das16,hojo18}. 
In fact, oxygen defects may modify significantly the structural and functional properties of perovskite 
thin films via changes in the $M$ oxidation states and their coupling with the lattice strain 
\cite{cazorla17b,lee17}. However, a number of theoretical works based on first-principles 
methods have agreed in that the combined action of $V_{O}$ and lattice strain may not affect
considerably the magnetic properties of Bi-based multiferroics \cite{ederer05,tsymbal12,chen12}. 

Here, we present new theoretical evidence showing that the presence of $V_{O}$ may in fact change 
radically the magnetic properties of multiferroic thin films via previously overlooked electro-structural 
mechanisms. We select BiCoO$_{3}$ (BCO) as the model multiferroic in which to perform first-principles 
calculations based on density functional theory (DFT) because (i)~this material already exhibits a 
large axial ratio in the absence of any strain, and (ii)~the magnetic effects that we predict  
can be realized on substrates that are commonly employed for growth of epitaxial oxide perovskite thin films 
(Fig.\ref{fig1}c). In particular, it is found that oxygen vacancies occupying specific lattice positions 
can induce the stabilization of full antiferromagnetic (AFM-G) super-tetragonal and ferrimagnetic (FiM) 
monoclinic polar phases, depending on the lattice strain. As a consequence, phase competition is enriched 
and magnetic functionalities further enhanced in comparison to perfectly stoichiometric thin films. 
We show that most of the results obtained in BCO thin films can be generalized to BFO and other Bi-based
multiferroics, hence our conclusions are of broad applicability and significance to the field of
functional materials.

\begin{figure*}
\centerline
        {\includegraphics[width=1.00\linewidth]{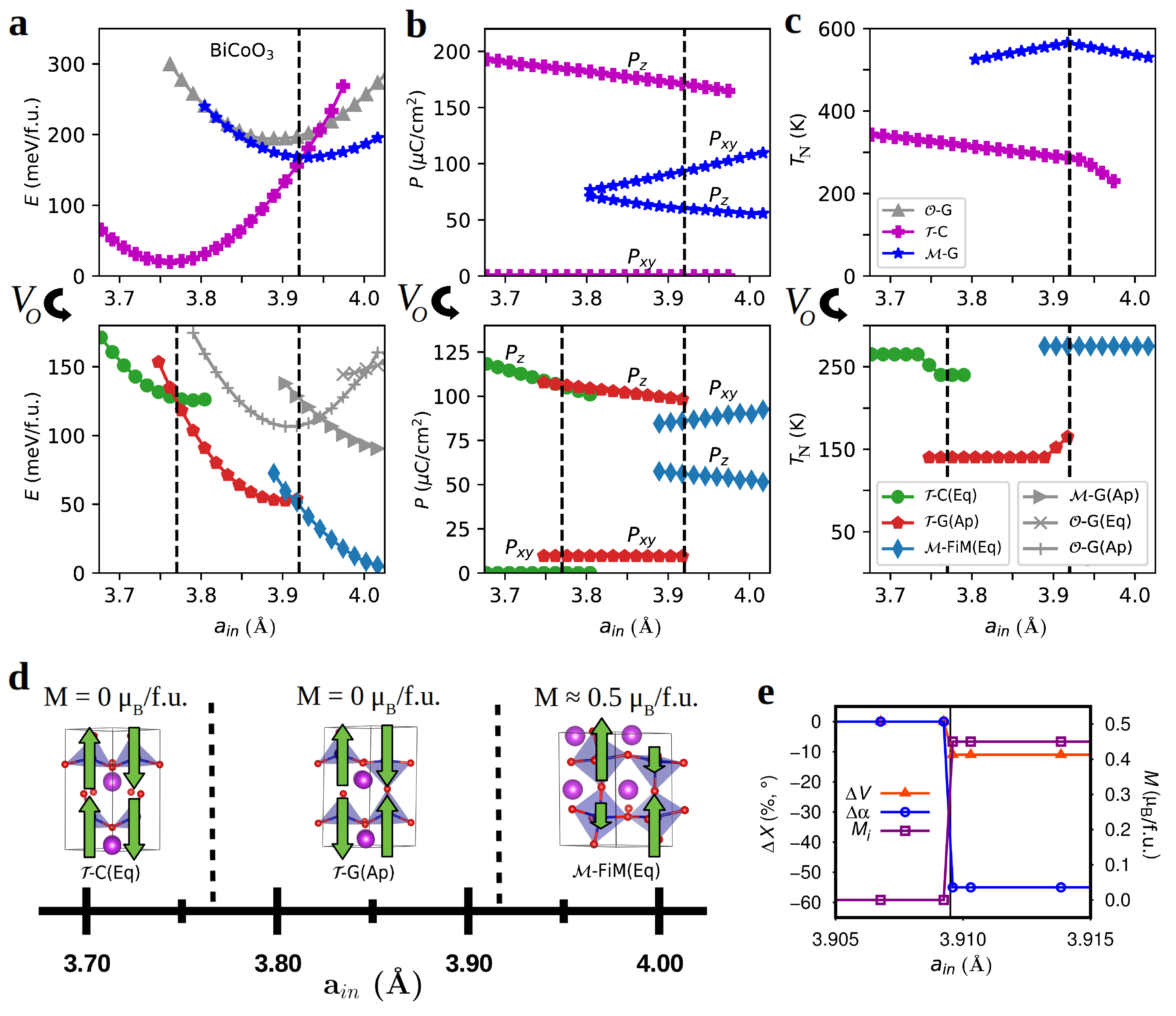}}
	\caption{Effects of $V_{O}$ on phase competition and functionality in BCO thin films.
	{\bf a} Zero-temperature energy of competing phases expressed as a function of in-plane lattice 
	parameter. Metastable phases are indicated by grey curves and strain-induced phase transitions 
	by vertical dashed lines; the oxygen vacancy positions leading to lowest energies, either 
	apical ``Ap'' or equatorial ``Eq'', are indicated within parentheses. ``G'' stands for G-type 
	antiferromagnetism, ``C'' for C-type antiferromagnetism, and ``FiM'' for ferrimagnetism. {\bf b} 
	Electric polarization of stoichiometric and non-stoichiometric ground-state phases. {\bf c} Magnetic 
	transition temperature of stoichiometric and non-stoichiometric ground-state phases. {\bf d} Phase 
	transition sequence occurring in non-stoichiometric BCO thin films under increasing $a_{in}$; the 
	green arrows represent the magnetic spin ordering in each phase. {\bf e} Change in volume, $\Delta V$, 
	change in electric polarization orientation, $\Delta \alpha$, and change in magnetic moment per 
	formula unit, $M_{i}$, associated to the multiferroic ${\cal T}$-G~$\to$~${\cal M}$-FiM phase transition.}
\label{fig2}
\end{figure*}

\section*{RESULTS}
{\bf $V_{O}$-induced effects on phase competition and functionality.}~Bulk BCO
presents a polar tetragonal ${\cal T}$ phase with a large axial ratio of $c/a 
\approx 1.3$ and relatively small lattice paramater $a = 3.76$~\AA~ \cite{cazorla18,cazorla17} 
(Fig.\ref{fig1}a). The competing structures are a non-polar orthorhombic ${\cal O}$ 
phase and a polar monoclinic ${\cal M}$ phase (Supplementary Fig.1); both competing 
phases have cells that are slightly distorted versions of the ideal cubic perovskite 
structure with $c/a \approx 1$. The polar phases in BCO present spontaneous polarizations 
along quite different crystallographic directions, namely, pseudocubic $[001]_{\rm pc}$ 
in ${\cal T}$ and $\sim[111]_{\rm pc}$ in ${\cal M}$. As regards magnetism, the ${\cal O}$ 
and ${\cal M}$ phases exhibit G-type antiferromagnetism (AFM-G) with a quite high 
N{\'e}el temperature, $T_{\rm N} \approx 500$~K, whereas the ${\cal T}$ phase C-type 
antiferromagnetism (AFM-C) with a relatively low $T_{\rm N}$ of $\approx 310$~K. 
In stoichiometric BCO thin films, and by completely neglecting temperature effects, a 
multiferroic ${\cal T} \to {\cal M}$ phase transition involving large structural, polar, 
and magnetic changes occurs at in-plane parameter $a_{in} = 3.91$~\AA ~\cite{cazorla18} 
(Figs.\ref{fig2}a-c).

Figures~\ref{fig2}a-c show the influence of neutral oxygen vacancies (Methods), $V_{O}$, 
on the structural, ferroelectric, and magnetic properties of BCO thin films (the 
accompanying changes in atomic lattice positions and energy band gap are reported 
in Supplementary Tables~1-4 and Supplementary Fig.2). For the smallest in-plane lattice 
parameters, a ${\cal T}$-C phase (magnetic spin ordering is indicated along with the 
structure symmetry) containing oxygen vacancies in equatorial (Eq) positions (Fig.\ref{fig1}a) 
renders the lowest energy. The electric polarization and N{\'e}el temperature in 
${\cal T}$-C(Eq) are significantly lower than in the analogous stoichiometric phase, in 
particular, we estimate differences of $\Delta P \approx -75$~$\mu$C~cm$^{-2}$ and 
$\Delta T_{\rm N} \approx -50$~K for same in-plane parameters. At $a_{in} = 3.77$~\AA, 
an unusual magnetic phase transition from AFM-C to AFM-G spin ordering occurs along with the 
appearance of a small in-plane electric polarization ($P_{xy} \sim 10$~$\mu$C~cm$^{-2}$) 
and change in $V_{O}$ position symmetry. The N{\'e}el temperature in the ${\cal T}$-G(Ap) 
phase is lower than in ${\cal T}$-C(Eq) by approximately $50$~K. Furthermore, at $a_{in} 
\ge 3.91$~\AA~ the system adopts a monoclinic ferrimagnetic (FiM) phase with oxygen vacancies 
in equatorial positions, ${\cal M}$-FiM(Eq), and a considerable net magnetization of $\approx 
0.5$~$\mu_{B}$ per formula unit (Fig.\ref{fig2}d). The N{\'e}el temperature in the 
${\cal M}$-FiM(Eq) phase is larger than in ${\cal T}$-G(Ap) and remains close to room temperature 
almost independently of $a_{in}$. (We have checked that the choice of the DFT energy functional 
and related technical parameters do not have a significant effect on these conclusions 
--Methods, Supplementary Figs.3-5, and Supplementary Methods--.)  

\begin{figure*}
\centerline
        {\includegraphics[width=1.00\linewidth]{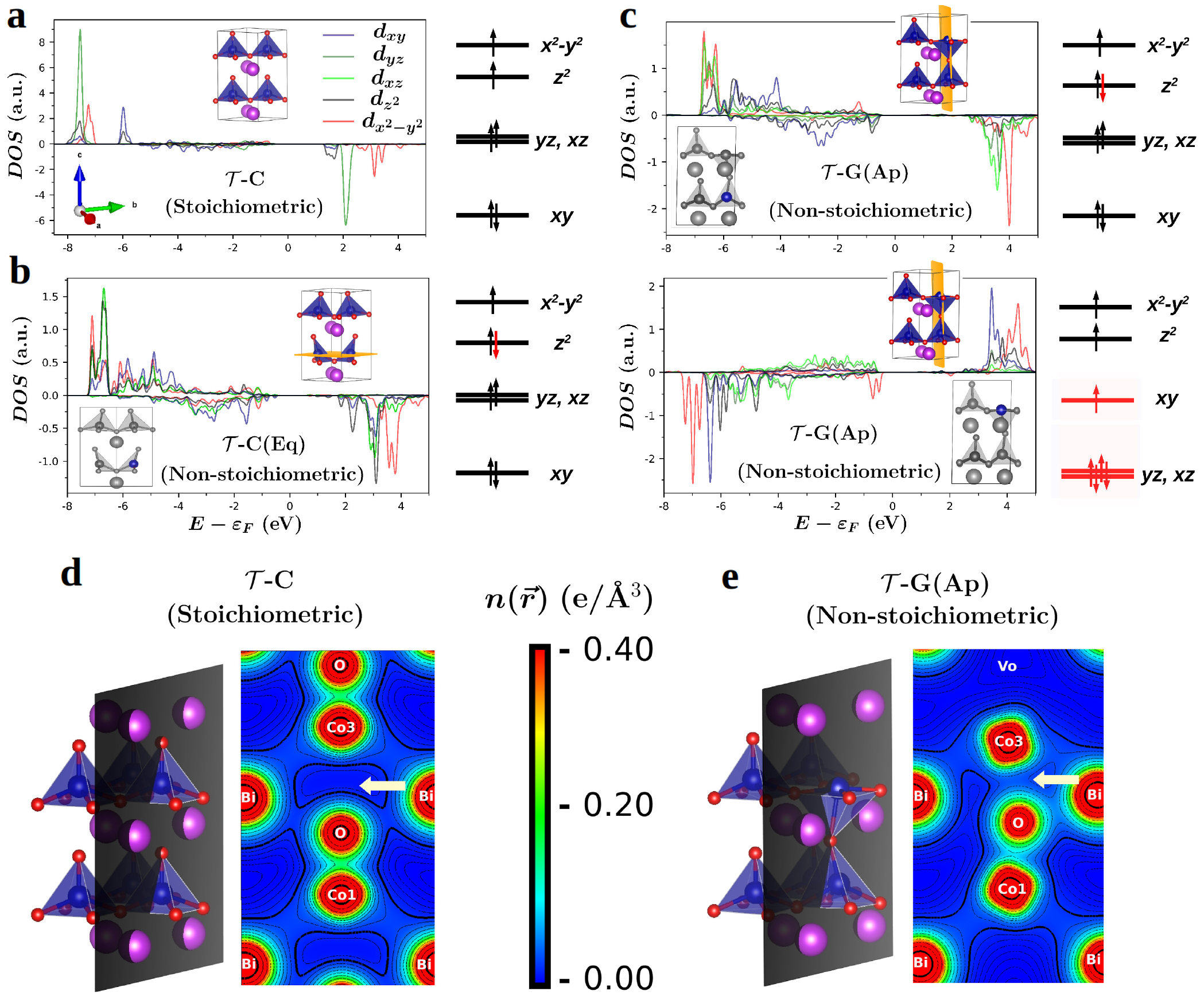}}
	\caption{Electronic, structural, and magnetic properties of ${\cal T}$ BCO thin films.
        {\bf a} Stoichiometric ${\cal T}$ phase.
	{\bf b} Non-stoichiometric ${\cal T}$ thin films with a stable $V_{O}$ in Eq position 
        and AFM-C spin ordering. The orange plane indicates the two Co ions that are reduced 
	as a consequence of creating a neutral Eq oxygen vacancy. The red arrow in the $d$-orbitals 
	occupation sketch indicates the difference with respect to the stoichiometric case.
	{\bf c} Non-stoichiometric ${\cal T}$ thin films with a stable $V_{O}$ in Ap position 
	and AFM-G spin ordering. The orange plane indicates the two Co ions that are reduced 
	as a consequence of creating a neutral Ap oxygen vacancy. 
        {\bf d} Charge density surface plot corresponding to the stoichiometric ${\cal T}$-C
	phase; the surface over which the charge density is calculated is indicated by a
	grey plane in the accompanying structural ball-stick representation. Isovalue paths 
	are represented	with black and coloured lines. {\bf e} Charge density surface plot 
	corresponding to the non-stoichiometric ${\cal T}$-G(Ap) phase. Regions of interest 
	describing Co--O bonds are indicated with yellow arrows.}
\label{fig3}
\end{figure*}

The physical mechanisms responsible for the two multiferroic phase transitions represented in
Fig.\ref{fig2}d will be explained in detail in the next subsections. Let us now comment
briefly on the functionality enhancement deriving from the ${\cal T}$-G(Ap)~$\to$~${\cal M}$-FiM(Eq)
transformation by keeping in mind that the non-stoichiometric ${\cal M}$ phase is ferrigmagnetic
and polar hence responsive to both external magnetic and electric fields. First, a large change 
in the electric polarization orientation involving a $\approx 60^{\circ}$ rotation is observed
during the transition (Fig.\ref{fig2}e); as a consequence, and in analogy to what has been observed 
in Pb(Zr$_{1-x}$Ti$_{x}$)O$_{3}$ alloys \cite{jaffe54,bellaiche00} and Bi(Fe$_{1-x}$Co$_{x}$)O$_{3}$ 
thin films \cite{shimizu16}, it should be possible to realize large piezoelectric responses under 
small electric bias at $a_{in} \approx 3.91$~\AA. Second, the sizeable changes in electric polarization 
and total magnetization in principle should allow for control of the polarization with magnetic fields 
and vice versa, which hints at the likely existence of large magnetoelectric couplings \cite{spalding07,spalding19}. 
And third, the out-of-plane lattice parameter shrinks by an impressive $\approx 11$\% (Fig.\ref{fig2}e) 
hence there is the possibility of realizing giant piezomagnetic responses \cite{domann17,nan17} and 
multicaloric effects \cite{stern18,cazorla18} through the application of external bias. In a more speculative 
vein, the change in $V_{O}$ position symmetry from Eq to Ap could lead to novel ionic transport phenomena 
driven by external magnetic, rather than electric, fields \cite{waskaas99}. As we will show later, similar 
magnetic phenomena are likely to occur also in other Bi-based multiferroic thin films, including 
BiFeO$_{3}$.  
\\

{\bf $V_{O}$-induced magnetic super-exchange interactions in the ${\cal T}$ phase.}~Figure~\ref{fig3}
summarizes the electronic, structural, and magnetic properties of stoichiometric and non-stoichiometric 
${\cal T}$ BCO thin films. In the stoichiometric ${\cal T}$-C phase (Fig.\ref{fig3}a), the square-pyramidal 
O$_{5}$ crystal field splits the electronic Co $d$ levels into nondegenerate $b_{2g}$ ($d_{xy}$), doubly 
degenerate $e_{g}$ ($d_{xz}$, $d_{yz}$), and nondegenerate $a_{1g}$ ($d_{z^{2}}$) and $b_{1g}$ ($d_{x^{2}-y^{2}}$). 
Our first-principles calculations render a high-spin Co state characterised by the electronic occupation 
$b^{2}_{2g}e^{2}_{g}a^{1}_{1g}b^{1}_{1g}$ and atomic spin moment $3.1$~$\mu_{B}$, in good agreement with 
the available experimental data \cite{oka10}. In the non-stoichiometric ${\cal T}$-C(Eq) phase (Fig.\ref{fig3}b), 
the splitting of electronic $d$ levels remains invariant with respect to the stoichiometric case and 
the occupation in the two cobalt ions nearest to the neutral $V_{O}$, which become reduced and are 
electronically equivalent, changes slightly to $b^{2}_{2g}e^{2}_{g}a^{2}_{1g}b^{1}_{1g}$ (depending on 
the choice of the technical DFT parameters this electronic distribution may vary somewhat --Supplementary
Fig.5 and Supplementary Methods--). 

Interestingly, when $V_{O}$ is created in an apical position and for specific $a_{in}$'s the magnetic spin 
ordering in the ${\cal T}$ phase changes to AFM-G. Figure~\ref{fig3}c shows the electronic density of states 
of the two reduced cobalt ions in the ${\cal T}$-G(Ap) phase, which in this case turn out to be electronically 
inequivalent. In particular, the doubly degenerate $e_{g}$ ($d_{xz}$, $d_{yz}$) orbitals in the cobalt ion 
closest to the apical $V_{O}$ (Co3, as labelled in Figs.\ref{fig3}d,e) undergo a significant energy reduction 
and become fully populated rendering the occupation state $e^{4}_{g}b^{1}_{2g}a^{1}_{1g}b^{1}_{1g}$, while the 
other reduced metal ion (Co1, as labelled in Figs.\ref{fig3}d,e) exhibits the more usual distribution 
$b^{2}_{2g}e^{2}_{g}a^{2}_{1g}b^{1}_{1g}$. These drastic electronic rearrangements are correlated with the 
appearance of a strong structural distortion in the system that pushes Co3 towards the oxygen atom underneath 
of it (see inversion of the corresponding O$_{5}$ square-pyramid in Figs.\ref{fig3}c,e) and tends to restore 
(partially) the missing magnetic super-exchange interactions along the out-of-plane direction. This super-exchange 
restoration mechanism, which is accompanied by an increase in covalency of the out-of-plane Co1--O--Co3 bonds 
and eventually leads to the stabilization of AFM-G spin ordering, is clearly imaged by plots of the electronic 
density in the plane containing Co1 and Co3 and oriented perpendicular to the substrate (Figs.\ref{fig3}d,e and 
yellow arrows therein). 

\begin{figure*}
\centerline
        {\includegraphics[width=1.00\linewidth]{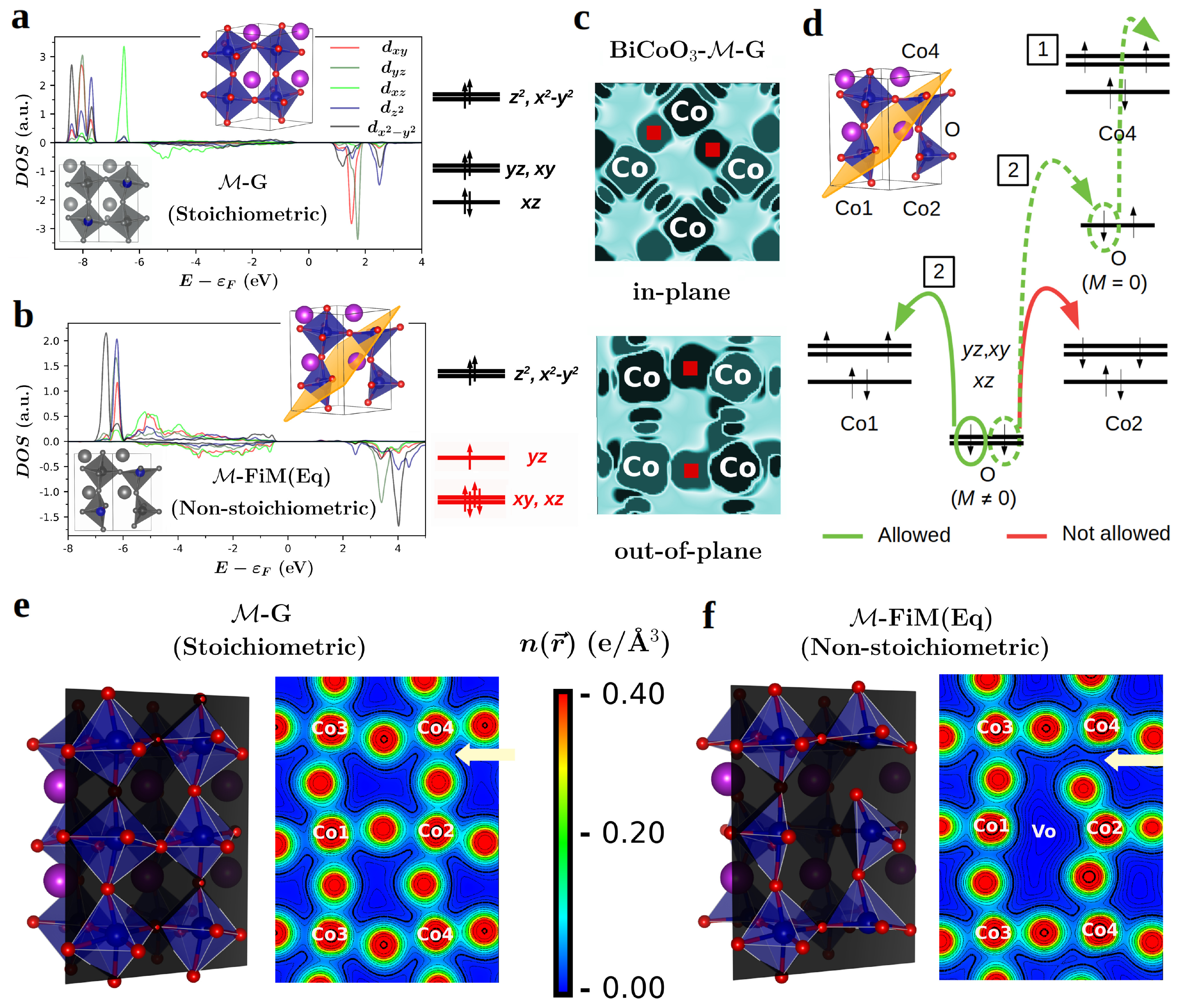}}
        \caption{Electronic, structural, and magnetic properties of ${\cal M}$ BCO thin films.
        {\bf a} Stoichiometric ${\cal M}$-G phase.
        {\bf b} Non-stoichiometric ${\cal M}$ thin films with a stable $V_{O}$ in Eq position 
	and FiM spin ordering. The orange plane indicates the two equivalent Co ions that are 
	reduced as a consequence of creating a neutral Eq oxygen vacancy. The red arrows in the 
	$d$-orbitals occupation sketch indicate the difference with respect to the stoichiometric case.
	{\bf c} Spin-up (dark green) and spin-down (light green) electronic charge densities
	calculated in stoichiometric ${\cal M}$ BCO thin films considering two perpendicular 
	planes; highly magnetized oxygen atoms are indicated with red squares.
	{\bf d} Inferred electronic hoppings enabling the stabilization of FiM spin ordering 
	in the ${\cal M}$-FiM(Eq) phase; numbers indicate the two events that are 
	likely to occur in a concerted manner; $d$ orbitals $a_{1g}$ and $b_{1g}$ are 
	disregarded due to their higher energies.
	{\bf e} Charge density surface plot corresponding to the stoichiometric ${\cal M}$-G
        phase; the surface over which the charge density is calculated is indicated by a
        grey plane in the accompanying structural ball-stick representation. Isovalue paths
        are represented with black and coloured lines.
	{\bf f} Charge density surface plot corresponding to the non-stoichiometric 
	${\cal M}$-FiM(Eq) phase. Regions of interest describing structural distortions and
	Co--O bonds are indicated with yellow arrows.}
\label{fig4}
\end{figure*}

\begin{figure*}
\centerline
        {\includegraphics[width=1.00\linewidth]{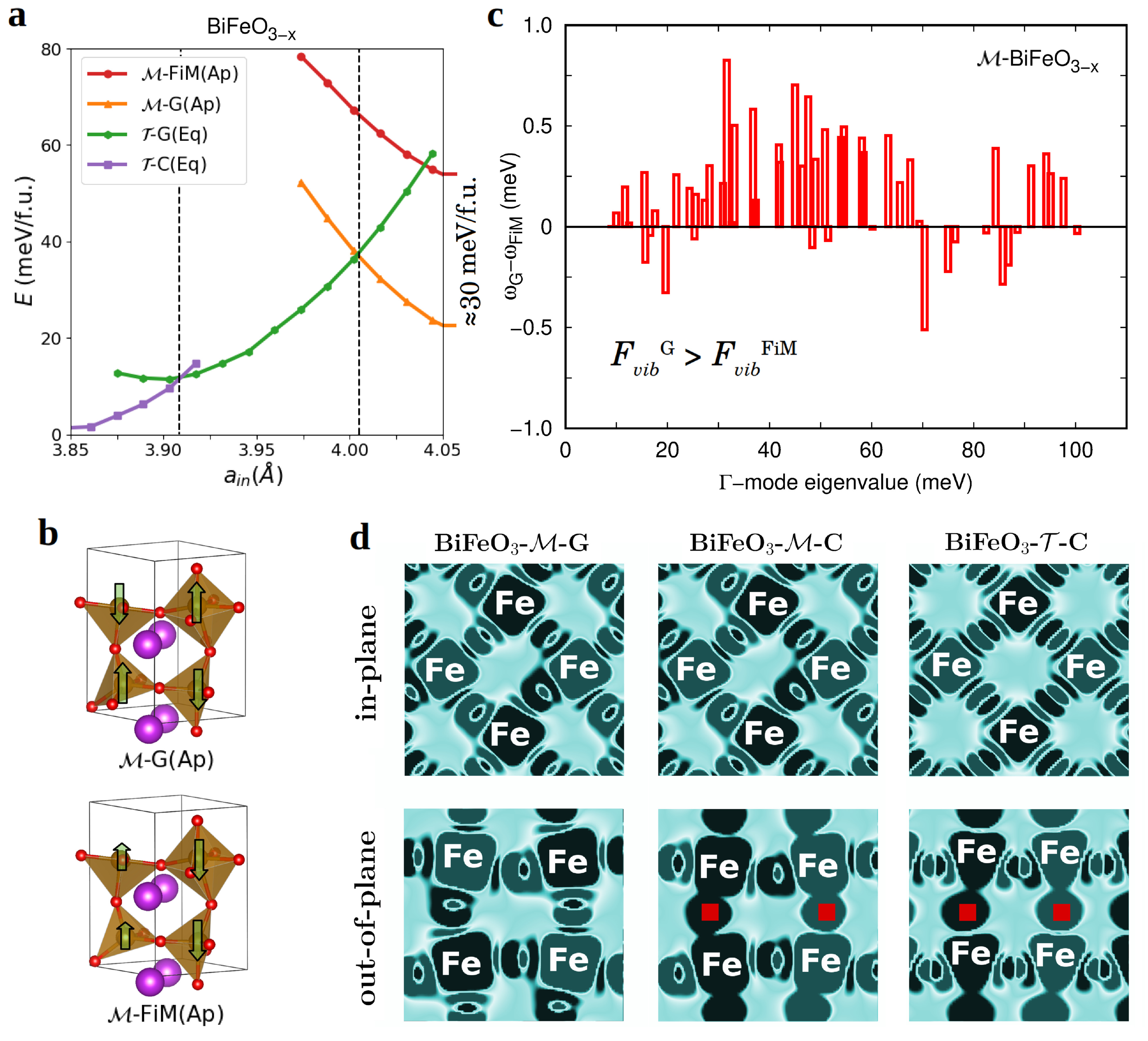}}
	\caption{Magnetic, structural, and vibrational properties of BFO thin films containing $V_{O}$. 
        {\bf a} Zero-temperature energy of competing phases expressed as a function of in-plane
        lattice parameter. {\bf b} Structural sketch of non-stoichiometric ${\cal M}$ phases considering
        different magnetic orderings. {\bf c} $\Gamma$-point phonon spectrum calculated in non-stoichiometric
        monoclinic BFO thin films; phonon frequencies in the ${\cal M}$-G(Ap) phase are higher in average
        than in ${\cal M}$-FiM(Ap) hence the vibrational free energy in the latter phase is lower. {\bf d} Spin-up
        (dark green) and spin-down (light green) electronic charge densities calculated in several stoichiometric
        BCO thin films considering two perpendicular planes; highly magnetized oxygen atoms are indicated 
	with red squares.}
\label{fig5}
\end{figure*}

The $V_{O}$-induced AFM-C~$\to$~AFM-G phase transition disclosed in large axial ratio BCO thin films
may shed some light on uncomprehended experimental observations of antiferromagnetic spin ordering
in other super-tetragonal multiferroic phases. For instance, in ${\cal T}$ BFO thin films several first-principles
works have predicted AFM-C spin ordering \cite{hatt10,dieguez11,heo17} whereas most experimental studies
indicate that AFM-G dominates \cite{bea09,dougall12}. As it will be explicitly shown later, by considering
the presence of oxygen vacancies in BFO thin films those theoretical and experimental results may be reconciled.
\\

{\bf $V_{O}$-induced stabilization of a ferrimagnetic ${\cal M}$ phase.}~Figure~\ref{fig4}
summarizes the electronic, structural, and magnetic properties of stoichiometric and non-stoichiometric
${\cal M}$ BCO thin films. In the stoichiometric ${\cal M}$-G phase (Fig.\ref{fig4}a), the octahedral 
O$_{6}$ crystal field splits the electronic Co $d$ levels into doubly degenerate $e_{g}$ ($d_{z^{2}}$,
$d_{x^{2}-y^{2}}$) and triply degenerate $t_{2g}$ ($d_{xy}$, $d_{yz}$, $d_{xz}$); a strong Jahn-Teller
distortion rendering a large $Q_{2}$ value of $0.37$~\AA~ \cite{cazorla16} lifts further the degeneracy 
in the $e_{g}$ and $t_{2g}$ manifolds \cite{halcrow13} and promotes the electronic occupation state 
$b^{2}_{2g}e^{2}_{g}b^{1}_{1g}a^{1}_{1g}$ (Fig.\ref{fig4}a). Remarkably, when specific $V_{O}$'s are 
created in equatorial positions (see next paragraph) the lowest-energy magnetic spin ordering changes 
to FiM and the net magnetization per formula unit in the ${\cal M}$-FiM(Eq) phase amounts to $\approx 
0.5$~$\mu_{B}$. The two Co ions that are reduced by the neutral vacancy present same magnetic moment 
orientation, same electronic occupancy $e^{4}_{g}b^{1}_{2g}b^{1}_{1g}a^{1}_{1g}$, and sit within the 
$[111]_{\rm pc}$ plane (Fig.\ref{fig4}b). 

How is possible that the two reduced Co ions are located along the diagonal of the pseudo-cubic unit 
cell rather than within the equatorial plane (that is, closest to the neutral $V_{O}$, in which case the
total magnetization would be null)? The ground-state ${\cal M}$-FiM(Eq) phase appears only when highly 
magnetized oxygen atoms ($0.1$--$0.2$~$\mu_{B}$) occupying equatorial positions in the stoichiometric ${\cal M}$-G 
crystal are removed (Fig.\ref{fig4}c). In that case, as we sketch in Fig.\ref{fig4}d, is not possible to 
reduce two neighbouring Co ions sitting in the equatorial plane (Co1 and Co2) due to Pauli exclusion 
principle. Consequently, pairs of metal ions with same magnetic moment and orientation (Co1 and Co4 in 
Fig.\ref{fig4}d) become reduced. For the couple of distant Co1 and Co4 ions to change their oxidation 
state and magnetic moment, however, the crystal needs to undergo sizable structural distortions involving 
the Bi and O atoms surrounding $V_{O}$ (Supplementary Tables~3-4). In particular, the non-magnetic oxygen atom 
in apical position just above Co2 acts as a bridge between the equatorial oxygen vacancy and Co4, by lending 
one of its electrons to the metal ion, hence reducing it, and receiving one electron from $V_{O}$ (Fig.\ref{fig4}d). 
This concerted electronic hopping mechanism can be inferred from plots of the electronic density in the plane 
containing Co1, Co2, Co3, and Co4 and oriented perpendicular to the substrate (Figs.\ref{fig4}e,f and yellow 
arrows therein). As can be appreciated therein, the covalency of the Co4--O bond is significantly reduced 
in the ${\cal M}$-FiM(Eq) phase as compared to the stoichiometric case, and a vertical tilt of the Co2--O 
bond that brings the apical oxygen closer to the equatorial $V_{O}$ is also evidenced.

The discovery of FiM spin ordering in the non-stoichiometric ${\cal M}$ phase motivated us to search for
similar magnetic states, even if metastable, in the other BCO thin film geometries ${\cal T}$ and ${\cal O}$.
The presence of highly magnetized O atoms was acknowledged in both stoichiometric phases (Supplementary
Fig.6), however upon removal of those oxygens we did not observe the appearance of any net magnetic moment
(neglecting small spin canting effects). These results confirm the importance of the electro-structural
mechanisms just described on facilitating the stabilization of FiM spin ordering. For instance, in the
${\cal T}$ phase the highly magnetized O atoms appear in apical positions (Supplementary Fig.6) hence 
the non-magnetic oxygens, which occupy equatorial positions and are somewhat clamped to the substrate, 
cannot act as electronic bridges between distant $V_{O}$'s and Co's. Meanwhile, in the ${\cal O}$ phase, 
which arguably is quite similar to ${\cal M}$ in terms of structure and magnetism (Supplementary Fig.1) \cite{cazorla18,cazorla17}, 
the lack of polar order and high dielectric permittivity makes the enabling Bi-O structural distortions 
(Supplementary Table~4) to be too high in energy; consequently, FiM spin ordering is frustrated. Based 
on these outcomes, we propose that the following three conditions are necessary for the stabilization of 
$V_{O}$-induced FiM spin ordering in magnetic A$M$O$_{3}$ oxide perovskites: (1)~lack of inversion symmetry 
leading to polar order and structural deformation ease, (2)~moderate axial ratio structures with $c/a 
\approx 1$ allowing for out-of-plane concerted electronic hoppings, and (3)~the existence of highly 
magnetized oxygen ions.

\section*{DISCUSSION}
The results obtained in BCO thin films raise the natural question: are there any other multiferroic 
materials in which similar $V_{O}$-induced magnetic phenomena may occur? To answer this question we 
investigated the special case of BiFeO$_{3}$ (BFO) and other Bi-based multiferroic (BiMnO$_{3}$ and 
BiCrO$_{3}$) thin films. Figure~\ref{fig5} encloses the energy, magnetic, and vibrational properties 
of non-stoichiometric BFO thin films. In order to be consistent with the notation employed heretofore, 
we label the usual rhombohedral-like monoclinic phase as ${\cal M}$ and the large axial ratio tetragonal-like 
phase as ${\cal T}$ (in spite of the fact that the space groups corresponding to those structures are 
different in BFO and BCO) \cite{heo17,cazorla15}. Our zero-temperature calculations (Fig.\ref{fig5}a) 
predict a ground-state ${\cal M}$-G(Ap) phase at $a_{in} \ge 4.01$~\AA, followed by ${\cal T}$-G(Eq) 
at $3.91 \le a_{in} \le 4.01$~\AA, and ${\cal T}$-C(Eq) at $a_{in} \le 3.91$~\AA. Indeed, when the likely 
existence of $V_{O}$ is explicitly considered in the simulations a broad $a_{in}$ region appears in which 
AFM-G spin ordering is stable in the ${\cal T}$ phase (that is missing in the corresponding stoichiometric 
system \cite{hatt10,heo17}). The causes of the stabilization of the ${\cal T}$-G(Ap) phase in BFO are 
very similar to those explained previously for ${\cal T}$-G(Ap) in BCO thin films (Fig.\ref{fig3}; we 
note that Ap $V_{O}$'s under tensile strain are in some ways equivalent to Eq $V_{O}$'s under compressive 
strain). As mentioned earlier, these results may shed new light on the origins of some unresolved 
discrepancies between theory and experiments regarding the determination of antiferromagnetic ordering 
in ${\cal T}$ BFO thin films \cite{hatt10,dieguez11,heo17,bea09,dougall12}. 

By creating an apical $V_{O}$ in the stoichiometric ${\cal M}$-C phase, we found a thus far neglected 
ferrimagnetic phase in monoclinic BFO thin films, ${\cal M}$-FiM(Ap). In this case the net magnetization 
per formula unit amounts also to $\approx 0.5$~$\mu_{B}$. Nevertheless, the ${\cal M}$-FiM(Ap) phase is 
metastable at zero temperature due to a small energy difference of $25-30$~meV per formula unit with 
respect to the ground-state phase (Fig.\ref{fig5}a). The atomic structure of the metastable ${\cal M}$-FiM(Ap) 
and ground-state ${\cal M}$-G(Ap) phases are highly distorted and surprisingly very similar 
(Fig.\ref{fig5}b and Supplementary Tables~5-6). Analysis of the $\Gamma$-point phonon modes (Fig.\ref{fig5}c), 
however, indicates that the ${\cal M}$-FiM(Ap) phase is vibrationally softer than ${\cal M}$-G(Ap). 
Consequently, the ${\cal M}$-FiM(Ap) phase may be entropically stabilized over ${\cal M}$-G(Ap) under 
increasing temperature since the zero-temperature energy difference between the two states is relatively 
small and the vibrational free energy of the former phase is more favourable \cite{cazorla18,cazorla17}. 
Our theoretical results, therefore, can be interpreted as evidence showing that the observation of 
``ferromagnetic'' behaviour in BFO thin films \cite{wang03} may be caused by the presence of oxygen 
vacancies, just as it has been suggested by other authors \cite{scott05}.

Figure~\ref{fig5}d shows the spin-up spin-down charge densities calculated in non-stoichiometric 
${\cal M}$ and ${\cal T}$ BFO thin films considering in-plane and out-of-plane surfaces. The reason 
why we could find just one FiM solution through the generation of neutral $V_{O}$'s in stoichiometric 
${\cal M}$ phases is now clear [see requirements (1)--(3) listed at the end of the previous section]: 
all oxygen atoms in the ${\cal M}$-G phase are non-magnetic whereas all apical O in the ${\cal M}$-C 
phase are highly magnetized. The stoichiometric ${\cal T}$-C phase also displays highly magnetized 
oxygen atoms in apical positions however, as we have explained before, FiM spin ordering hardly 
can be generated in geometries exhibiting $c/a \gg 1$. Furthermore, we repeated the same type of 
calculations and analysis in monoclinic-like BiMnO$_{3}$ and BiCrO$_{3}$ thin films, which fulfill 
conditions (1) and (2) explained above, and found that the appearance of FiM spin ordering is also 
correlated with the presence of highly magnetized oxygen atoms (Supplementary Fig.7). In particular, 
a ${\cal M}$-FiM(Ap) phase displaying a net magnetization of $\approx 0.2$~$\mu_{B}$ is found in BiCrO$_{3}$. 
These results confirm that the simple rules provided in this work for prediction of FiM phases in 
non-stoichiometric multiferroic thin films are robust and general.  
\\

In summary, by using first-principles calculations we have disclosed a number of previously overlooked
electro-structural mechanisms induced by the presence of oxygen vacancies that facilitate the stabilization 
of unexpected magnetic states in multiferroic thin films. In particular, AFM-G and FiM spin orderings 
may naturally appear in large axial ratio and monoclinic phases under certain lattice strain conditions. 
Our theoretical results may clarify the origins of some long-standing controversies in BFO, the paradigm 
of single-phase multiferroics and one of the most intensively studied functional materials. We provide 
general and simple rules to fundamentally understand and systematically predict FiM phases in non-stoichiometric 
multiferroics, thus offering new approaches for the rational engineering of bettered functional materials. 
The present work shows that oxygen vacancies should be considered as a design opportunity to create new 
funcionalities, especially as related to magnetism, in multiferroic thin films.

\section*{METHODS}
{\bf Density functional theory calculations.}~First-principles spin-polarized calculations based on density 
functional theory (DFT) are performed with the generalized gradient approximation proposed by Perdew, Burke 
and Ernzerhof (GGA-PBE) as implemented in the VASP package \cite{vasp,pbe96}. We employ the ``Hubbard-U'' 
scheme derived by Dudarev \textit{et al.} to deal with the $3d$ electrons in Co~(Fe) atoms and, as done in 
previous works, a $U$ value of $6$~($4$)~eV is adopted \cite{cazorla18,cazorla13,cazorla17}. We use the 
``projected augmented wave'' method \cite{bloch94} considering the following electronic states as valence: 
Co's $4s^{1}3d^{8}$, Fe's $3p^{6}4s^{1}3d^{7}$, Mn's $4s^{1}3d^{6}$, Cr's $4s^{1}3d^{5}$, Bi's $6s^{2}5d^{10}6p^{3}$, 
and O's $2s^{2}2p^{4}$. The energy cut-off is truncated at $650$~eV and we employ a $\Gamma$-centered ${\bf k}$-point 
grid of $4 \times 6 \times 6$ for a $2 \times \sqrt{2} \times \sqrt{2}$ supercell containing $20$ atoms (that is, 
four formula units) \cite{cazorla15}. Periodic boundary conditions are applied along the three lattice-vector
directions. Thin film geometry relaxations are carried out by using a conjugated gradient algorithm that allows 
to change the simulation-cell volume and atomic positions while constraining the length and orientation of the 
two in-plane lattice vectors. The geometry relaxations are stopped once the forces acting on the ions are 
smaller than $0.01$~eV/\AA. We have checked the vibrational stability of every phase by estimating the lattice 
phonons at the $\Gamma$-point with the small-displacement method \cite{boronat17} and considering central differences 
for the calculation of atomic forces derivatives.  

Non-stoichiometric phases are generated by removing one oxygen atom from an apical or equatorial position in 
the $20$-atoms simulation cell, thus rendering the chemical composition BiCoO$_{3-x}$ with $x = 0.25$. Apical and 
equatorial $V_{O}$ positions are investigated systematically by generating all inequivalent configurations in 
all competing phases and considering all possible magnetic spin orderings (FM, AFM-G, AFM-G, and AFM-A \cite{cazorla17}). 
The results presented in the main text are obtained by assuming neutral oxygen vacancies since we have found 
that neutral $V_{O}$'s are energetically more favourable than charged oxygen vacancies (Supplementary Fig.8,
Supplementary Table~7, and Supplementary Methods). In particular, we have employed the following well-established 
formula to estimate the ranking of $V_{O}$ formation energies as a function of charge, $q$, \cite{pasquarello12}: 
\begin{eqnarray}
	E_{def}[V_{O}^q] & = & E[V_{O}^q] - E_{stoi} + E^{q}_{corr} - n_{O}\mu_{O} + \nonumber \\
	                 &   & q \left[ \epsilon_{F} + \epsilon_{v} + \Delta V \right]~,
\label{eq:charged}
\end{eqnarray}	
where $E[V_{O}^q]$ is the energy of the non-stoichiometric system containing the oxygen vacancy, $E_{stoi}$ the energy
of the corresponding stoichiometric system, $E^{q}_{corr}$ a finite-size supercell correction, $n_{O}$ the number
of created $V_{O}$'s (typically equal to $1$ in our calculations), $\mu_{O}$ the chemical potential of oxygen atoms, 
$\epsilon_{F}$ the Fermi energy in the non-stoichiometric system, $\epsilon_{v}$ the top energy in the valence band 
of the non-stoichiometric system, and $\Delta V$ a term used for aligning the electrostatic potentials of the stoichiometric 
and defective supercells. In order to calculate $E^{q}_{corr}$ and $\Delta V$, we have followed the methods explained in 
work \cite{pasquarello12}. According to our estimations, neutral $V_{O}$'s ($q = 0$) are energetically more favourable 
than charged vacancies ($q = +2$~$e$) by about $\sim 1$~eV per formula unit (Supplementary Fig.8, Supplementary Table~7, 
and Supplementary Methods).

We have performed several tests to assess the influence of the adopted DFT exchange-correlation functional and
$U$ value on our theoretical predictions (Supplementary Figs.3-5 and Supplementary Methods). Specifically, we repeated 
most calculations by considering the PBEsol functional \cite{pbesol} and $2 \le U \le 6$~eV values. It is found that 
the main conclusions presented in the main text are not affected qualitatively by the choice of the $U$ parameter
or exchange-correlation functional. At the quantitative level, the $a_{in}$ parameters at which the phase transitions 
occur and the electronic occupations that are deduced from electronic density plots change a little in some cases 
(Supplementary Figs.3-5 and Supplementary Methods); however, the structural properties and energy ranking of the 
competing phases estimated in most $a_{in}$ cases remain invariant. For a detailed discussion on these technical aspects, 
see Supplementary Methods.

Regarding the estimation of the electric polarization, $P$, we started by employing the Berry phase formalism \cite{vanderbilt93}. 
However, the presence of oxygen vacancies induces a notable reduction in the energy band gap of the system 
(Supplementary Fig.2) that in some cases frustrates the determination of the corresponding Berry phase (due to 
the appearance of intermediate metallic phases). In order to overcome such a limitation, we opted for calculating 
the electric polarization perturbatively. Specifically, we estimate $P$ with the formula \cite{cazorla15}:  
\begin{equation} 
P_{\alpha} = \frac{1}{\Omega} \sum_{\kappa\beta}  Z_{\kappa\beta \alpha}^{*} u_{\kappa\beta}~, 
\label{eq:polarization}
\end{equation} 
where $\Omega$ is the volume of the cell, $\kappa$ runs over all the atoms, $\alpha,\beta = x, y, z$ 
represent Cartesian directions, $\bf{u}_{\kappa}$ is the displacement vector of the $\kappa$-th atom as 
referred to a non-polar reference phase, and $\boldsymbol{Z}^{*}_{\kappa}$ the Born effective charge tensor 
calculated in the non-polar reference state. In the stoichiometric phases we do not find the technical 
limitations just explained for non-stoichiometric systems, hence in that case we have been able to compare 
the $P$ values obtained with the Berry phase approach (exact) and Eq.(\ref{eq:polarization}) (approximate). 
According to our estimations, the electric polarizations calculated perturbatively are accurate to within 
$\sim 10$\% of, and systematically larger than, the $P$ values calculated with the Berry phase method. It 
is reasonable to assume then a similar level of accuracy in the $P$ values estimated in non-stoichiometric 
thin films that are reported in Fig.\ref{fig2}.  
\\

{\bf Heisenberg model Monte Carlo simulations.}~To simulate the effects of thermal excitations on 
magnetic ordering in ${\cal T}$, ${\cal O}$, and ${\cal M}$ BCO thin films, we construct several 
spin Heisenberg models of the form $\hat{H} = \frac{1}{2} \sum_{ij} J^{(0)}_{ij} S_{i}S_{j}$, in 
which the value of the involved exchange constants are obtained from zero-temperature DFT calculations 
(see works \cite{cazorla18,cazorla13,cazorla17} for the technical details on the determination of the 
$J^{(0)}_{ij}$ parameters). We use those models to perform Monte Carlo (MC) simulations in a 
periodically-repeated simulation box of $20 \times 20 \times 20$ spins; thermal averages are computed 
from runs of $50,000$ MC sweeps after equilibration. These simulations allow us to monitor the $T$-dependence 
of the magnetic ordering through the computation of the AFM-C (in the ${\cal T}$ phase) and AFM-G (in 
the ${\cal O}$ and ${\cal M}$ phases) order parameters, namely, $S^{\rm C} \equiv \frac{1}{N} \sum_{i} 
(-1)^{n_{ix}+n_{iy}} S_{iz}$ and $S^{\rm G} \equiv \frac{1}{N} \sum_{i} (-1)^{n_{ix}+n_{iy}+n_{iz}} S_{iz}$. 
Here, $n_{ix}$, $n_{iy}$, and $n_{iz}$ are the three integers locating the $i$-th lattice cell, and $N$ 
is the total number of spins in the simulation box. For the calculation of $S^{\rm C}$ and $S^{\rm G}$, 
we considered only the $z$ component of the spins because a small symmetry-breaking magnetic anisotropy 
was introduced in the Hamiltonian in order to facilitate the numerical analysis \cite{cazorla18,cazorla13,cazorla17}.

\section*{DATA AVAILABILITY}
The data that support the findings of this study are available from the corresponding author (C.C.) upon
reasonable request.

\section*{ACKNOWLEDGMENTS}
Computational resources and technical assistance were provided by the Australian Government
and the Government of Western Australia through the National Computational Infrastructure
(NCI) and Magnus under the National Computational Merit Allocation Scheme and The Pawsey
Supercomputing Centre.\\

\section*{AUTHOR CONTRIBUTIONS}
C.C. conceived the study and planned the research. C.M. and C.C. performed the theoretical
calculations. Results were discussed by C.M., D.C., and C.C. The manuscript was written by 
C.M., D.C., and C.C.\\

\section*{ADDITIONAL INFORMATION}
Supplementary information is available in the online version of the paper.\\

\section*{COMPETING INTERESTS}
The authors declare no competing interests.

\end{document}